\documentclass[twocolumn,showpacs,preprintnumbers,amsmath,amssymb]{revtex4}
\usepackage{graphicx}

\begin{document}
\title{Comment on ``Quantum Strategy Without Entanglement''}
\author{Shengchao Ding}
\email{dingshengchao@ict.ac.cn} \affiliation{Institute of Computing Technology, Chinese
Academy of Sciences, Beijing 100080, P.R.China\\
Graduate University of the Chinese Academy of Sciences, Beijing 100080, P.R.China\\}

\date{\today}

\begin{abstract}
We make remarks on the paper of Du et al (quant-ph/0011078) by pointing out that the
quantum strategy proposed by the paper is trivial to the card game and proposing a simple
classical strategy to make the game in classical sense fair too.
\end{abstract}


\maketitle

In the paper of Du et al~\cite{Du00CardGame}, they present a two players' game just as
the following card game. There are three cards, otherwise identical, except for the
following markings: the first card has a circle on each side; the second card has a dot
on each side; the third card has a circle on one side and a dot on the other. Alice puts
the three cards in a black box and shakes it to randomize the three cards. Bob is allowed
to draw one card out from the box without seeing the cards. If the card has the same
marks on both sides, Alice wins one. Otherwise, Bob wins one. Obviously, this game is
unfair to Bob because that the possibility to draw out an identical face card in three
cards is $\frac{2}{3}$, and Alice has the expected payoff $\bar{\pi}_A =
\frac{2}{3}\times 1 + \frac{1}{3}\times (-1) = \frac{1}{3}$ while Bob has the expected
payoff $\bar{\pi}_B = \frac{1}{3}\times 1 + \frac{2}{3}\times (-1) = -\frac{1}{3}$
~\cite{Grabbe05QuantumGame}. Actually, if Bob see all the upper faces of the three cards
before drawing, he will make sure that the card with different upper face in three cards
must be a card with the identical faces, so he will randomly draw one of the two other
cards, which make him win the game with a fifty-fifty chance. But this observation is
forbidden.

Du et al propose a quantum strategy to make this game fair by adding two principles such
as:
\begin{enumerate}
\item \label{principle:1}Allow Bob to make a single \textit{query} by calling a quantum oracle to the black box;
\item \label{principle:2}Allow Bob to withdraw from the game once he knows the upper face of the card he
draws is different to the upper faces of two other cards.
\end{enumerate}

It is obvious that only adding the principle \ref{principle:2} to the card game helps Bob
nothing because that he isn't able to make such a decision without any other information.
So the principle \ref{principle:1} is necessary. However, we will show that the quantum
oracle proposed by Du et al is trivial.

\begin{figure}
    \includegraphics[width=0.42\textwidth]{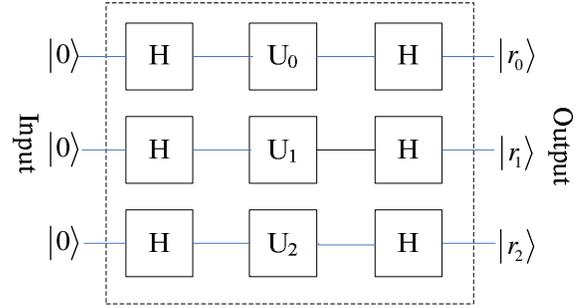}
    \caption{The proposed quantum oracle for card game}\label{fig:original}
\end{figure}

Considering the quantum oracle, let the card's state be $|0\rangle$ if the upper face is
a circle or $|1\rangle$ if the upper face is a dot. So the three-card state is $|r\rangle
= |r_0r_1r_2\rangle$, where $r_k\in\{0,1\}$. The proposed quantum oracle in
\cite{Du00CardGame} is just as Fig.~\ref{fig:original} which is little different from the
original figure in their paper. Note that the inner structure of their oracle is shown
here.
\begin{figure}
    \includegraphics[width=0.25\textwidth]{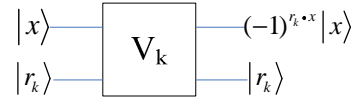}
    \caption{The equivalent quantum circuit of $U_k$}\label{fig:equivalent}
\end{figure}

In Fig.~\ref{fig:original}, as part of the quantum oracle, the following unitary matrix
is required:
\begin{equation}
U_k = \left(\begin{array}{cc}
              1 & 0 \\
              0 & e^{i\pi r_k}
            \end{array}\right)
            =\left\{
               \begin{array}{ll}
                 I_2, & r_k=0; \\
                 \sigma_z, & r_k=1.
               \end{array}
             \right.
\end{equation}

If we want to construct the oracle we should know all values of $r_k$($k=0,1,2$).
Actually, $U_k$ is just the quantum circuit shown in Fig.~\ref{fig:equivalent}, i.e.
$V_k|x\rangle|r_k\rangle = (-1)^{r_k\cdot x}|x\rangle|r_k\rangle$. So the original oracle
is equivalent to be added one input, i.e. $|r_k\rangle$, for each $|0\rangle$. The
transformation is as follows.
\begin{equation}
\begin{array}{cl}
   & (H\otimes I)V_k(H\otimes I)|0\rangle|r_k\rangle \\
  = & (H\otimes I)V_k(\frac{|0\rangle+|1\rangle}{\sqrt{2}})|r_k\rangle \\
  = & (H\otimes I)\frac{1}{\sqrt{2}}\left( (-1)^{r_k\cdot 0}|0\rangle + (-1)^{r_k\cdot 1}|1\rangle \right) |r_k\rangle \\
  = & \frac{1}{2}\left( (|0\rangle+|1\rangle) + (-1)^{r_k}(|0\rangle - |1\rangle) \right)|r_k\rangle \\
  = & \frac{1}{2}\left(( 1+(-1)^{r_k})|0\rangle + (1-(-1)^{r_k})|1\rangle \right)|r_k\rangle \\
  = & |r_k\rangle |r_k\rangle
\end{array}
\end{equation}

Obviously this transformation is trivial to the game. So adding the
principle~\ref{principle:1} to the game is equivalent to adding a third player from which
Bob could know the information about all the $r_k$. If Bob is allowed to know the
information about $r_k$ and the principle~\ref{principle:2} is available, the game in
classical sense is fair to Bob too.

In the classical sense, in fact, we can simply alert a principle of the game and make it
fair to both. The strategy is that if Alice wins she only gets one but if Bob wins he
gets two. Now the payoffs are $\bar{\pi}_A = \frac{2}{3}\times 1+\frac{1}{3}\times
 (-2)=0$, $\bar{\pi}_B = \frac{1}{3}\times 2 + \frac{2}{3}\times(-1)=0$, and the game is
 a zero-sum game thus a fair game.


\begin{thebibliography}{99}
\bibitem{Du00CardGame} J. Du, X. Xu, H. Li, M. Shi, X. Zhou and R. Han, quant-ph/0011078.
\bibitem{Grabbe05QuantumGame} J.O. Grabbe, quant-ph/0506219.
\end{thebibliography}
\end{document}